\documentclass[12pt, a4paper]{article}

\usepackage{graphicx}

\usepackage{stmaryrd}
\usepackage{overcite}
\usepackage{verbatim}
\usepackage{achemso}
\usepackage{verbatim}

\usepackage{multirow}
\usepackage{amsmath}
\usepackage{amsfonts}
\usepackage{amsgen}
\usepackage{longtable}
\usepackage{lscape}
\usepackage{threeparttable}
\usepackage{supertabular}
\usepackage{multicol}
\usepackage{float}
\usepackage{multirow}
\usepackage{setspace}
\usepackage{rotating}
\usepackage{amssymb}
\usepackage{layout}
\usepackage{wrapfig}
\usepackage{subfig}

\begin{document}

\begin{center}

{\bf\large{Double Potential Step Chronoamperometry at a Microband Electrode: Theory and Experiment}}

\hspace{2cm}

{\bf\large Edward O. Barnes, Linhongjia Xiong, Kristopher R. Ward and Richard G. Compton*}

*Corresponding author

Department~of~Chemistry, Physical~and~Theoretical~Chemistry~Laboratory, Oxford~University,
South~Parks~Road, Oxford, OX1~3QZ, United~Kingdom.
Fax:~+44~(0)~1865~275410; Tel:~+44~(0)~1865~275413.
Email:~richard.compton@chem.ox.ac.uk
\vspace{1cm}

NOTICE: this is the author's version of a work that was accepted for publication in \emph{The Journal of Electroanalytical Chemistry}. Changes resulting from the publishing process, such as peer review, editing, corrections, structural formatting, and other quality control mechanisms may not be reflected in this document. Changes may have been made to this work since it was submitted for publication. A definitive version was subsequently published in \emph{The Journal of Electroanalytical Chemistry}, DOI 10.1016/j.jelechem.2013.05.002.
\end{center}

\clearpage

\section*{Abstract}

Numerical simulation is used to characterise double potential step chronoamperometry at a microband electrode for a simple redox process, A $+$ e$^-$ $\rightleftharpoons$ B, under conditions of full support such that diffusion is the only active form of mass transport. The method is shown to be highly sensitive for the measurement of the diffusion coefficients of both A and B, and is applied to the one electron oxidation of decamethylferrocene (DMFc), DMFc $-$ e$^-$ $\rightleftharpoons$ DMFc$^{+}$, in the room temperature ionic liquid 1-propyl-3-methylimidazolium bistrifluoromethylsulfonylimide. Theory and experiment are seen to be in excellent agreement and the following values of the diffusion coefficients were measured at 298 K: $D_\mathrm{DMFc} = 2.50 \times 10^{-7}$ cm$^2$ s$^{-1}$ and $D_\mathrm{DMFc^{+}} = 9.50 \times 10^{-8}$ cm$^2$ s$^{-1}$.

\section*{Keywords}

Microband electrode; Double potential step, Diffusion coefficients, measurement of; Chronoamperometry; Numerical simulation.

\clearpage

\section{Introduction}

The field of electrochemistry has been transformed by the introduction of microelectrodes\cite{Compton2010}, which have the properties of enhanced mass transport allowing steady states to be achieved\cite{Aoki1981}, operate with reduced Ohmic drop\cite{Amatore2008a} and often allow a two electrode setup to be employed with a combined reference/counter electrode\cite{Barnes2010a}. The most widely used microelectrode is the microdisc, which is easily fabricated and has well characterised properties\cite{Oldham1981, Heinze1981}.

The application of microdisc electrodes are many and varied, including using single potential step chronoamperometry to determine the diffusion coefficient of a species of interest, $D_\mathrm{A}$, and simultaneously either its concentration, $c_\mathrm{A}$, or the number of electrons transfered, $n$, provided one of these two parameters is known\cite{Paddon2007}. In a single step chronoamperometry experiment, the potential applied to the working electrode is stepped from a value where no reaction occurs, to one where reduction (or oxidation) occurs at a mass transport controlled rate. The well established Shoup-Szabo equation then describes the current measured within 0.6 \% error\cite{Shoup1982}:
\begin{equation}
i = 4nFc_\mathrm{A}D_\mathrm{A}r_ef(\tau)
\end{equation}
where
\begin{equation}
f(\tau) = 0.7854 + 0.4432\tau^{-0.5} + 0.2146\text{exp}\left(-0.3912\tau^{-0.5}\right)
\end{equation}
and
\begin{equation}
\tau = \frac{D_\mathrm{A}t}{r_e^2}
\end{equation}
where $i$ is the measured current (A), $F$ is the Faraday constant and $r_e$ is the radius of the microelectrode (m). The change in the current's dependency on $D_\mathrm{A}$ (from $\sqrt{D_\mathrm{A}}$ at short times to linearly dependent at long times), and the direct proportionality of the current to $nc_\mathrm{A}$ at all times, allows both $D_\mathrm{A}$ and the product $nc_\mathrm{A}$ to be determined by fitting of the Shoup-Szabo equation to experimental data. A knowledge of either $n$ or $c_\mathrm{A}$ then allows calculation of the final unknown.

Using double potential step chronoamperometry, in conjunction with  numerical simulation, this method can be extended to find not only $D_\mathrm{A}$, but also $D_\mathrm{B}$, the diffusion coefficient of the other member of the redox couple\cite{Klymenko2004}. In these experiments, the potential is first stepped in the same manner as for single potential step experiments, and held at this reducing potential for a set time, $t_s$, before being stepped a second time to a value where the reverse reaction occurs at a mass transport controlled rate, and species B is converted back to species A. Numerical simulations of double potential step experiments can then be used, with $D_\mathrm{A}$, $c_\mathrm{A}$ and $D_\mathrm{B}$ as input parameters, and these parameters optimised to obtain a best fit.

In addition to these simple yet powerful techniques, microelectrodes of various geometries have found uses in a wide variety of applications, including weakly supported voltammetry\cite{Streeter2008b} (where their very low Ohmic drop is of great advantage in the study of the effects of migration), generator/collector systems\cite{Barnes2012}, electrochemical sensors\cite{Bai2013, Huan2012, Metters2012}, and studies of ionic liquid properties\cite{Walsh2010, Broder2007}.

Despite these extensive uses of microelectrodes, investigations of double potential step chronoamperometry have largely confined themselves to microdiscs, except for one case each at spherical\cite{Galvez1992} and hemispherical\cite{Limon-Petersen2009} electrodes; we can find no report at microband electrodes. This study therefore extends the theory of double potential step chronoamperometry to microband electrodes. Note these electrodes are being manufactured by Nanoflux PTE LTD\textregistered\cite{nanoflux} and MicruX\textregistered\cite{micrux}, have been noted for their ease of construction, cheapness and durability\cite{Welford2001}, and are finding increasing use in electrochemical sensors.

In addition to these advantages, microband electrodes have emerged as being experimentally useful, and have found steadily increasing use in hydrodynamic voltammetry in a flow cell\cite{Compton1993b, Alden1995, Alden1996a, Alden1996b, Ueno2003, Amatore2011a, Amatore2011, Matthews2012}, in dual electrode generator-collector mode\cite{Fosset1991, Unwin1991, Rajantie2001, Paixao2003, Svir2003, Amatore2006}, and in impedance spectroscopy\cite{Compton1995a, Alden1996c}. One of the key differences in electrochemistry carried out at microbands, as opposed to microdiscs, is the absence of a true steady state in chornoamperometry at microbands\cite{Aoki1987}. This has the effect that, at all times, the measured current is dependent on both $D$ and $\sqrt{D}$. This lack of a true steady state will make behavior observed in double potential step chronoamperometry at microbands different to that at microdiscs, and perhaps more useful.

The study of ionic liquids has received a huge amount of attention in recent years (see \cite{Barrosse-Antle2010} and \cite{Silvester2006a} for reviews). Being composed entirely of mobile ions, their intrinsic conductivity removes the need to add any supporting electrolyte to carry out electrochemical experiments. They also have a near zero volatility, and often have a very wide electrochemical window. This has led to their application in electrochemical gas sensors\cite{Xiong2012}, as well as sensors for temperature\cite{Xiong2012} and humidity\cite{Xiong2012a}. A further property of ionic liquids is that they can show drastic differences between the diffusion coefficients of members of a redox pair, \emph{e.g.} the factor of 30 difference between the diffusion coefficients of O$_2$ and O$_2^{\bullet -}$ in the ionic liquid hexyltriethylammonium bis-((trifluoromethyl)sulfonyl)imide ([N6222][N(Tf)2])\cite{Buzzeo2003}. This is due to the extremely strong ionic interactions hindering the movement of charged species. This large difference between diffusion coefficients in ionic liquids necessitates accurately measuring both of them if their electrochemistry is to be fully understood, and so performing double potential step chronoamperomtry in the solvents is a necessity.

In this paper, a model for the numerical simulation of double potential step chronoamperometry at a microband electrode is developed, and used to asses the extent to which this technique can be used to determine both $D_\mathrm{A}$ and $D_\mathrm{B}$ at a microband electrode. We validate this model by simulating the double potential step chronoamperometry of decamethylferrocene in the ionic liquid 1-propyl-3-methylimidazolium bistrifluoromethylsulfonylimide (PmimNTf$_2$) and fitting experimental data.

\section{Theory}

\begin{figure}[h]
\begin{center}
\includegraphics[width = 0.5\textwidth]{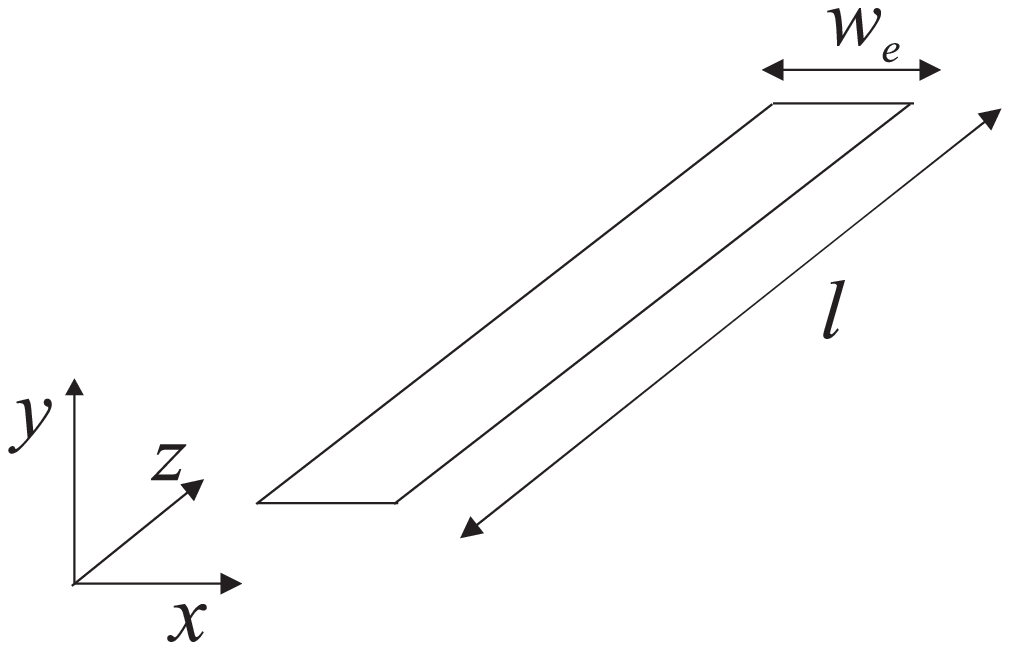}
\caption{Schematic diagram of a microband electrode.} \label{SCHEMATIC}
\end{center}
\end{figure}

In this paper, we develop a model to solve the problem of double potential step chronoamperometry, as discussed in the introduction, at a planar microband electrode. A schematic of the electrode being used is shown in Figure \ref{SCHEMATIC}, indicating the parameters $w_e$ (the electrode width), $l$ (the electrode length) and the orientation of the Cartesian coordinates. A simple one electron redox couple is considered:
\begin{equation}
\mathrm{A} \pm \mathrm{e^-} \rightleftharpoons \mathrm {B}
\end{equation}
If the microband is considered to be infinitely long in the $z$ direction (relative to its width in the $x$ direction, see Figure \ref{SCHEMATIC}), the edge effects at the ends of the electrode can be neglected, along with diffusion along the length of the electrode, reducing the mass transport equation to be solved to two dimensions:
\begin{equation}
\frac{\partial{c_\mathrm{i}}}{\partial{t}} = D_\mathrm{i}\left(\frac{\partial^2c_\mathrm{i}}{\partial{x^2}} + \frac{\partial^2c_\mathrm{i}}{\partial{y^2}}\right)
\end{equation}
All symbols are defined in Table \ref{DIMENSIONAL}.

Subject to appropriate boundary conditions described below, this equations is used to simulate the time evolution of the concentration of species in solution.

\subsection{Boundary Conditions}

\begin{figure}[h]
\begin{center}
\includegraphics[width = 0.9\textwidth]{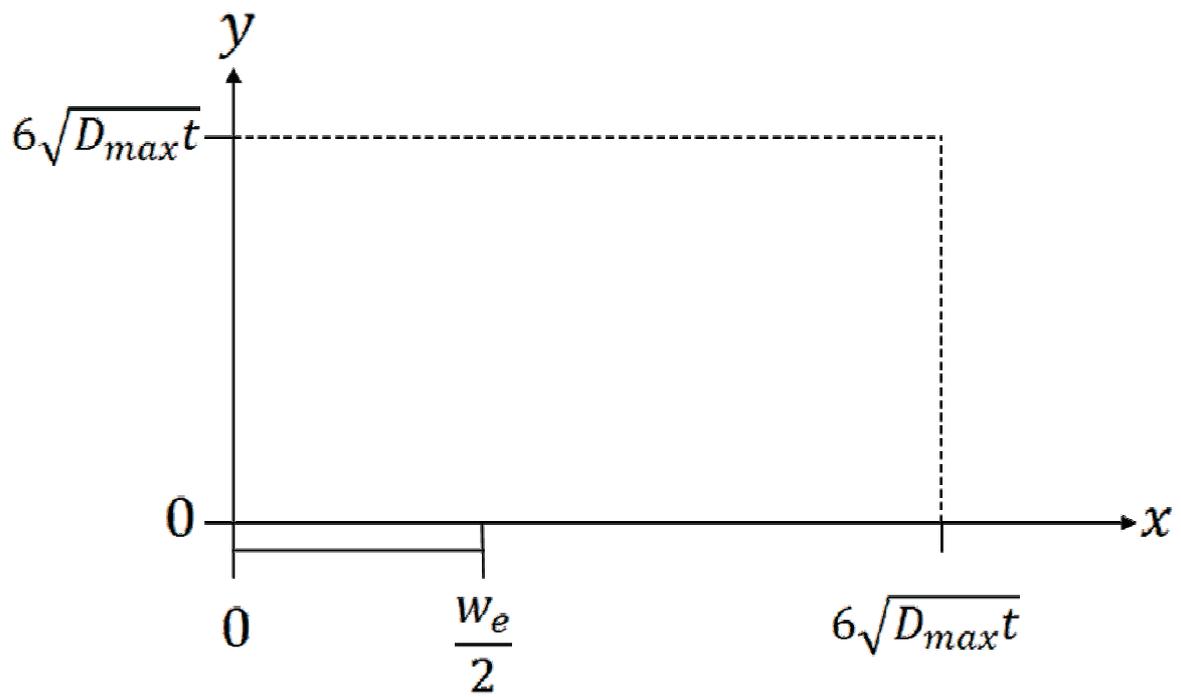}
\caption{Schematic diagram of the simulation space used in the model.} \label{SIM SPACE}
\end{center}
\end{figure}

The simulation space used is shown schematically in Figure \ref{SIM SPACE}. Due to symmetry around $x = 0$, we only need simulate half of the electrode and surrounding solution, with a zero flux condition imposed by symmetry at $x=0$:
\begin{equation}
\left(\frac{\partial{c_\mathrm{i}}}{\partial{x}}\right)_{x=0}=0
\end{equation}
A zero flux condition is imposed on the insulating surface around the electrode:
\begin{equation}
\left(\frac{\partial{c_\mathrm{i}}}{\partial{y}}\right)_{x > \frac{w_e}{2}} = 0
\end{equation}
Before the experiment begins at $t = 0$, the potential is set such that no reaction occurs and no current is drawn:
\begin{equation}
t < 0\text{; all } x \text{; all } y 
\begin{cases}
c_\mathrm{A} = c_\mathrm{A}^{*}\\
c_\mathrm{B} = 0
\end{cases}
\end{equation}
At times after $t=0$, at the electrode surface ($x < \frac{w_e}{2}$, $y=0$) the boundary conditions for each species depend on the applied potential. Before some switching time, $t_s$, the potential applied to the electrode is such that species A is reduced at a mass transport controlled rate to form species B. Hence, when $t < t_s$:
\begin{equation}
t < t_s\text{; } x < \frac{w_e}{2}\text{; } y = 0
\begin{cases}
c_\mathrm{A} = 0\\
D_\mathrm{B}\left(\frac{\partial{c_\mathrm{B}}}{\partial{y}}\right) = -D_\mathrm{A}\left(\frac{\partial{c_\mathrm{A}}}{\partial{y}}\right)
\end{cases}
\end{equation}
After $t_s$, the potential is stepped to a more positive value such that now species B is oxidised at a mass transport controlled rate back to species A:
\begin{equation}
t \geq t_s\text{; } x < \frac{w_e}{2}\text{; } y = 0
\begin{cases}
c_\mathrm{B} = 0\\
D_\mathrm{A}\left(\frac{\partial{c_\mathrm{A}}}{\partial{y}}\right) = -D_\mathrm{B}\left(\frac{\partial{c_\mathrm{B}}}{\partial{y}}\right)
\end{cases}
\end{equation}

The bulk solution boundaries were set at $6\sqrt{D_\mathrm{max}t_\mathrm{max}}$ from the electrode in both the $x$ and $y$ direction, where $D_\mathrm{max}$ and $t_\mathrm{max}$ are the maximum diffusion coefficient in the system and the total time of the experiment respectively\cite{Gavaghan1998b, Gavaghan1998a, Gavaghan1998}. At these boundaries, bulk concentrations are assumed:
\begin{equation}
x = \frac{w_e}{2}+6\sqrt{D_\mathrm{max}t_\mathrm{max}}\text{; } y = 6\sqrt{D_\mathrm{max}t_\mathrm{max}}\text{; } c_\mathrm{A} = c_\mathrm{A}^{*}\text{; } c_\mathrm{B} = 0
\end{equation}

\subsection{Dimensionless Parameters}

The model described above is normalised by introducing a series of dimensionless parameters, which reduces the number of variables and removes scaling factors. For example, concentrations are expressed relative the the bulk concentration of species A:
\begin{equation}
C_\mathrm{i} = \frac{c_\mathrm{i}}{c_\mathrm{A}^{*}}
\end{equation}
and dimensionless lengths expressed relative to the electrode width:
\begin{equation}
X = \frac{x}{w_e} \quad\quad Y = \frac{y}{w_e}
\end{equation}
A full list of normalised parameters and their definitions is given in Table \ref{DIMENSIONLESS}. Once normalised, the mass transport equation is given by:
\begin{equation}
\frac{\partial{C_\mathrm{i}}}{\partial{\tau}} = D^{'}_\mathrm{i}\left(\frac{\partial^2{C_\mathrm{i}}}{\partial{X^2}} + \frac{\partial^2{C_\mathrm{i}}}{\partial{Y^2}}\right)
\end{equation}
The normalised boundary conditions are listed in Table \ref{BOUNDARY CONDITIONS}.

To calculate the current, the flux must be evaluated at each point on the electrode, and summed up over the entire electrode surface. The total dimensionless flux, $j$, is given by:
\begin{equation}
j = 2\int^{X=0}_{X=0.5}{\left(\frac{\partial{C_\mathrm{A}}}{\partial{Y}}\right)_{Y=0}} \mathrm{d}X
\end{equation}
and the dimensional current given by:
\begin{equation}
I = -FAJ
\end{equation}
where $F$ is the Faraday constant, $A$ is the electrode area in m$^2$, and $J$ is the dimensional flux, given by:
\begin{equation}
J = \frac{c_\mathrm{A}^\mathrm{*}D_\mathrm{A}}{w_e}j
\end{equation}
giving:
\begin{equation}
I = -c_\mathrm{A}^{*}D_\mathrm{A}lFj
\end{equation}
assuming an initial reduction. If instead, species A is oxidised, the current is trivially multiplied by minus 1.

\subsection{Numerical Methods}

All equations are discretised according to the Crank-Nicolson method\cite{Crank1947} and solved over discrete spatial and temporal grids using the alternating direction implicit (ADI) method in conjunction with the Thomas algorithm for an n-diagonal matrix\cite{Press2007}. The form temporal grid employed is one which has been successfully used in previous models\cite{Amatore2003, Amatore2004,Barnes2013}. After a potential step, the temporal grid is initially uniform, but after some defined change time $\tau_c$ it expands:
\begin{eqnarray}
\tau \leq \tau_c & \quad & \tau_{k+1} = \tau_k + \Delta_\tau\tau\\
\tau_c < \tau < \tau_s & \quad & \tau_{k+1} = \tau_k + \gamma_\tau\left(\tau_k - \tau_{k-1}\right)
\end{eqnarray}
After the potential switch, the temporal grid repeats its form but offset by an amount $\tau_s$. It should be noted that since microband electrodes exhibit pseudo steady state behavior\cite{Aoki1987}, a denser temporal grid is required than for microdisc simulations where a steady state concentration profile allows for large time steps at high $\tau$.

\begin{figure}[h]
\begin{center}
\includegraphics[width = 0.9\textwidth]{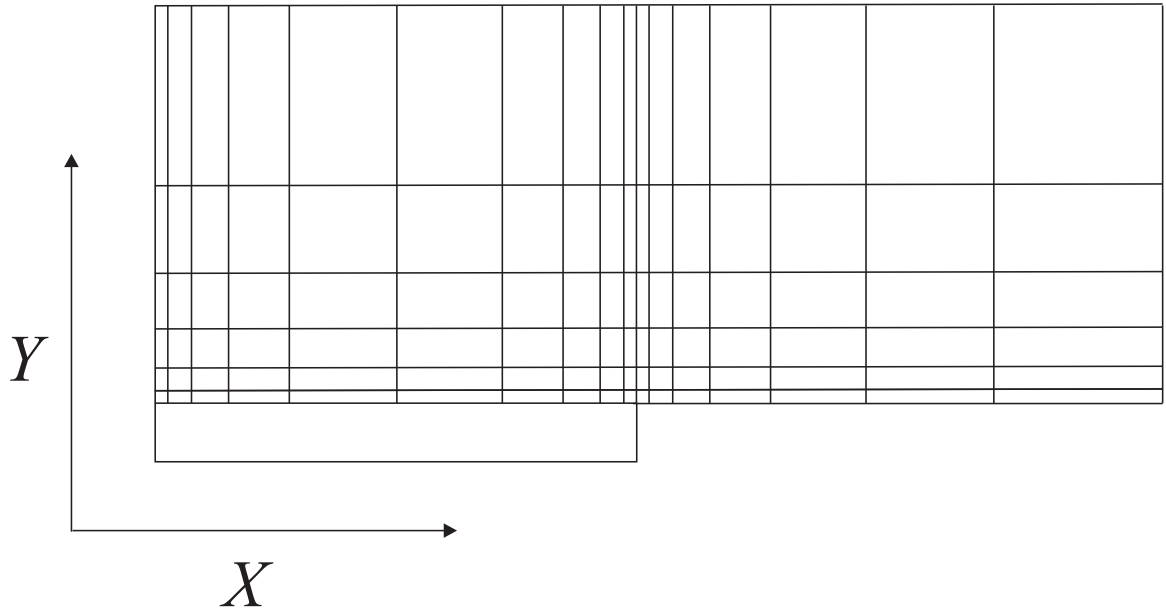}
\caption{Schematic diagram of the spatial grid employed (some lines removed for clarity).} \label{GRID}
\end{center}
\end{figure}

The spatial grid used is shown schematically in Figure \ref{GRID} (some lines removed for clarity). After an initial step of $\Delta_s$, the grid expands from $X = 0$ and from the electrode edge ($X = 0.5$) in both direction in a way directly analogous to the temporal grid, with an expansion coefficient of $\gamma_s$. Convergence studies found grid parameter values of: $\Delta_\tau = 1 \times 10^{-7}$, $\tau_c = 1 \times 10^{-4}$, $\gamma_\tau = 1.0001$, $\Delta_s = 1 \times 10^{-5}$ and $\gamma_s = 1.1$ were sufficient to give results such that making either grid ten times denser changed the simulation result by less than 0.5\%. The simulations were coded in C++ and run on an Intel(R) Xeon(R) 2.26 GHz PC with 2.25 GB RAM.

\subsection{Simulation of Cyclic Voltammetry}

When simulating cyclic voltammetry, rather than double potential step chronoamperometry, at the microband, changes must be made to the boundary conditions and temporal grid.

The boundary conditions at the electrode surface are now given by Butler-Volmer kinetics:
\begin{equation}
t > 0\text{; } x < \frac{w_e}{2}\text{; } y = 0
\begin{cases}
D_\mathrm{A}\frac{\partial{c_\mathrm{A}}}{\partial{y}} = k^0\left[c_\mathrm{A}^0\text{exp}\left(-\alpha\frac{\left(E - E_f\right)F}{RT}\right) - c_\mathrm{B}^0\text{exp}\left(\left(1-\alpha\right)\frac{\left(E - E_f\right)F}{RF}\right)\right]\\
D_\mathrm{B}\frac{\partial{c_\mathrm{B}}}{\partial{y}} = - D_\mathrm{A}\frac{\partial{c_\mathrm{A}}}{\partial{y}}
\end{cases}
\end{equation}
where $k$ is the electrochemical rate constant (m s$^{-1}$), $c_\mathrm{i}^0$ is the surface concentration of species i (mol m$^{-3}$), $E$ is the potential applied to the working electrode (V), $E_f$ is the formal potential of the A/B redox couple (V), and $\alpha$ is the transfer coefficient. In dimensionless parameters, this boundary condition becomes:
\begin{equation}
\tau > 0\text{; } X < 0.5\text{; } Y = 0
\begin{cases}
\frac{\partial{C_\mathrm{A}}}{\partial{Y}} = K^0\left[C_\mathrm{A}^0\text{exp}\left(-\alpha\theta\right) - C_\mathrm{B}^0\text{exp}\left(\left(1-\alpha\right)\theta\right)\right]\\
D^{'}_\mathrm{B}\frac{\partial{C_\mathrm{B}}}{\partial{Y}} = - \frac{\partial{C_\mathrm{A}}}{\partial{Y}}
\end{cases}
\end{equation}
The relationship between the applied potential, $E$, and time is dependent on the scan rate, $\nu$ (V s$^{-1}$):
\begin{equation}
E = |E_s - E_v - \nu t| + E_v
\end{equation}
where $E_s$ and $E_v$ are the start and vertex potentials respectively, or in dimensionless parameters:
\begin{equation}
\theta = |\theta_s - \theta_v - \sigma\tau | + \theta_v
\end{equation}
The same spatial grid is used as for chronoamperometry, but rather than an expanding temporal grid, a regular one is now used. A parameter $\theta_\mathrm{div}$ is defined as the number of temporal grid points per unit theta swept out. In this way, before the switching potential:
\begin{equation}
\theta_k = \theta_{k-1} - \frac{1}{\theta_\mathrm{div}}
\end{equation}
and after it:
\begin{equation}
\theta_k = \theta_{k-1} + \frac{1}{\theta_\mathrm{div}}
\end{equation}
A $\theta_\mathrm{div}$ value of 100 was found sufficient to converge the simulations.

\section{Theoretical Results}

In order to validate the program used in this study, the simulated dimensionless flux for the first step of a double potential step chronoamperometry experiment can be compared to the equation developed by Aoki \emph{et. al.}\cite{Aoki1987}:
\begin{equation}
j = \frac{1}{\sqrt{\pi\tau}} + 0.97 - 1.10\text{exp}\left(\frac{-9.90}{|\text{ln}\left(12.37\tau\right)|}\right)
\end{equation}

\begin{figure}[h]
\begin{center}
\includegraphics[width = 0.9\textwidth]{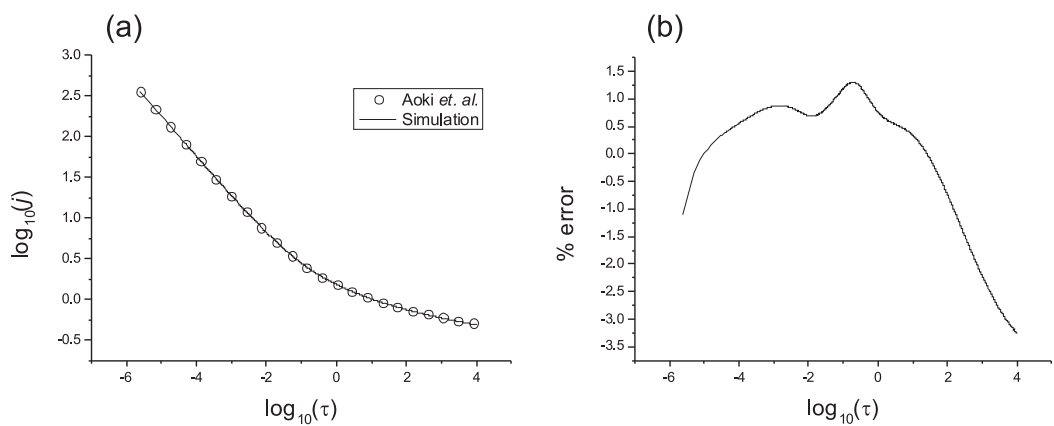}
\caption{(a) Dimensionless flux during the first step of a double potential step chronoamperometry experiment at a band electrode calculated via simulation (line) and Aoki's equation (circles). (b) Percentage error between the simulation used in this study and Aoki's equation.} \label{ERROR}
\end{center}
\end{figure}

The comparison between simulations from this model and from the analytical equation proposed by Aoki \emph{et. al.} is shown in Figure \ref{ERROR}. Excellent agreement is seen between the two.

As discussed in the introduction, double potential step chronoamperometry can be used to obtain values for the diffusion coefficients of both the oxidised and reduced form of the species under investigation. The model described above can be used to probe the extent to which microband electrodes are useful in this respect, and under what conditions values can be most easily obtained.

\begin{figure}[h]
\begin{center}
\includegraphics[width = 0.9\textwidth]{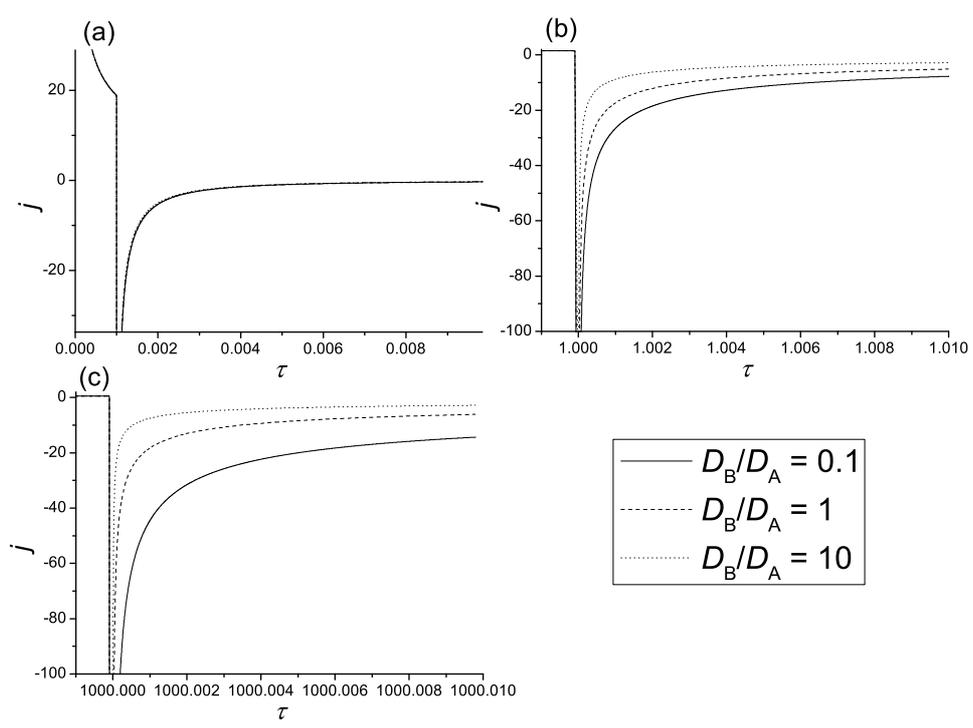}
\caption{Dimensionless flux at short times after the second potential step for $D^{'}_\mathrm{B}$ values of 0.1, 1 and 10, and $\tau_s$ values of (a) 0.001, (b) 1 and (c) 1000} \label{DIFFERENT D}
\end{center}
\end{figure}

Simulations of double potential step chronoamperometry were carried out with various values of $\tau_s$, the time at which the potential is switched, and $D^{'}_\mathrm{B}$. The results, zoomed in on the parts of the chronoamperograms of interest, are shown in Figure \ref{DIFFERENT D}. $\tau_s$ values of 0.001, 1 and 1000 were used, as well as $D^{'}_\mathrm{B}$ values of 0.1, 1 and 10. By comparing the results from different diffusion coefficients at various switching times, it can be seen whether switching the potential after a short time or a long time will give the greatest discrepancies between different diffusion coefficients, and hence allow the most accurate determinations of $D_\mathrm{B}$. In Figure \ref{DIFFERENT D}, it is immediately apparent that a large value of $\tau_s$ is desirable for discriminating between different values of $D^{'}_\mathrm{B}$. For small values of $\tau_s$, species B does not have time to diffuse away from the electrode to any significant extent, and the currents drawn after $\tau_s$ are essentially identical for all three values of $D^{'}_\mathrm{B}$. For larger values of $\tau_s$, however, species B does diffuse away from the electrode to different extents for different $D^{'}_\mathrm{B}$, therefore giving very different responses after the second potential step.

Tables \ref{TAUS=1} to \ref{TAUS=1000} tabulate reference values for the dimensionless flux, $j$, measured after the second potential step, for $\tau_s$ values of 1, 10, 100 and 1000, each with a wide range of $D^{'}_\mathrm{B}$ values. The (dimensional) value of $D_\mathrm{A}$ can be extracted by fitting experimental data to the first transient, then $D_\mathrm{B}$ estimated by comparing the second step transient with these reference tables. The pseudo-steady state nature of the diffusion to a microband allows higher resolution between different $D^{'}_\mathrm{B}$ values to be achieved with higher $\tau_s$ values, approaching a limit as $\tau_s$ approaches $\infty$. 

\section{Experimental}

\subsection{Chemicals}

Ferrocene (Fe(C$_5$H$_5$)$_2$, Aldrich, 98\%), acetonitrile (MeCN, Fischer Scientific, dried and distilled, 99\%), tetra-n-butylammonium perchlorate (TBAP, Fluka, Puriss electrochemical grade, 99\%), decamethylferrocene (Fe(C$_{10}$H$_{15}$)$_{2}$, Fluka, 95\%)and 1-propyl-3-methylimidazolium bistrifluoromethylsulfonylimide (PmimNTf$_2$, kindly donated by Queen's University, Belfast) were used as received without further purification.

\subsection{Instrumental} 

All electrochemical experiments were carried out using a computer-controlled PGSTAT30-Autolab potentiostat (Eco-Chemie, Netherlands). Solutions were housed in a sealed glass vial, with a three-electrode arrangement consisting of either a 5.0 $\mu$m radius Pt disc or a Pt microband (see below) working electrode, a silver wire reference electrode and Pt coil wire counter electrode. The platinum microdisc working electrode was polished on soft lapping pads (Kemet Ltd., UK) using alumina powders (Buehler, IL) of sizes 1.0, 0.3 and 0.05 $\mu$m. The electrode radius was calibrated electrochemically by analysing the steady-state voltammetry of a 2.0 mM solution of ferrocene in MeCN containing 0.1 M TBAP, using a diffusion coefficient for ferrocene of $2.30\times10^{-9}$ m$^2$ s$^{-1}$ at $298$ K\cite{Rogers2008a}. The ionic liquid was degassed under vacuum overnight to remove water and other impurities. All experiments were performed inside a thermostated box (previously described by Evans et al.\cite{Evans2004}) which also functioned as a Faraday cage. Unless specified, the temperature was maintained at 298 ($\pm{0.5}$) K. A platinum microband electrode of dimensions 6.28 mm $\times$ 50 $\mu$m was fabricated using the method of Wadhawan \emph{et. al}\cite{Welford2001}.

\section{Experimental Results}

To validate the band electrode model outlined above, we can compare experimental and theoretical data at a band electrode to data collected at a microdisc electrode, simulated using the well established model developed by Klymenko \emph{et. al}\cite{Klymenko2004}.

\subsection{Double Potential Step Chronoamperometry of Decamethylferrocene}

\begin{figure}[h]
\begin{center}
\includegraphics[width = 0.9\textwidth]{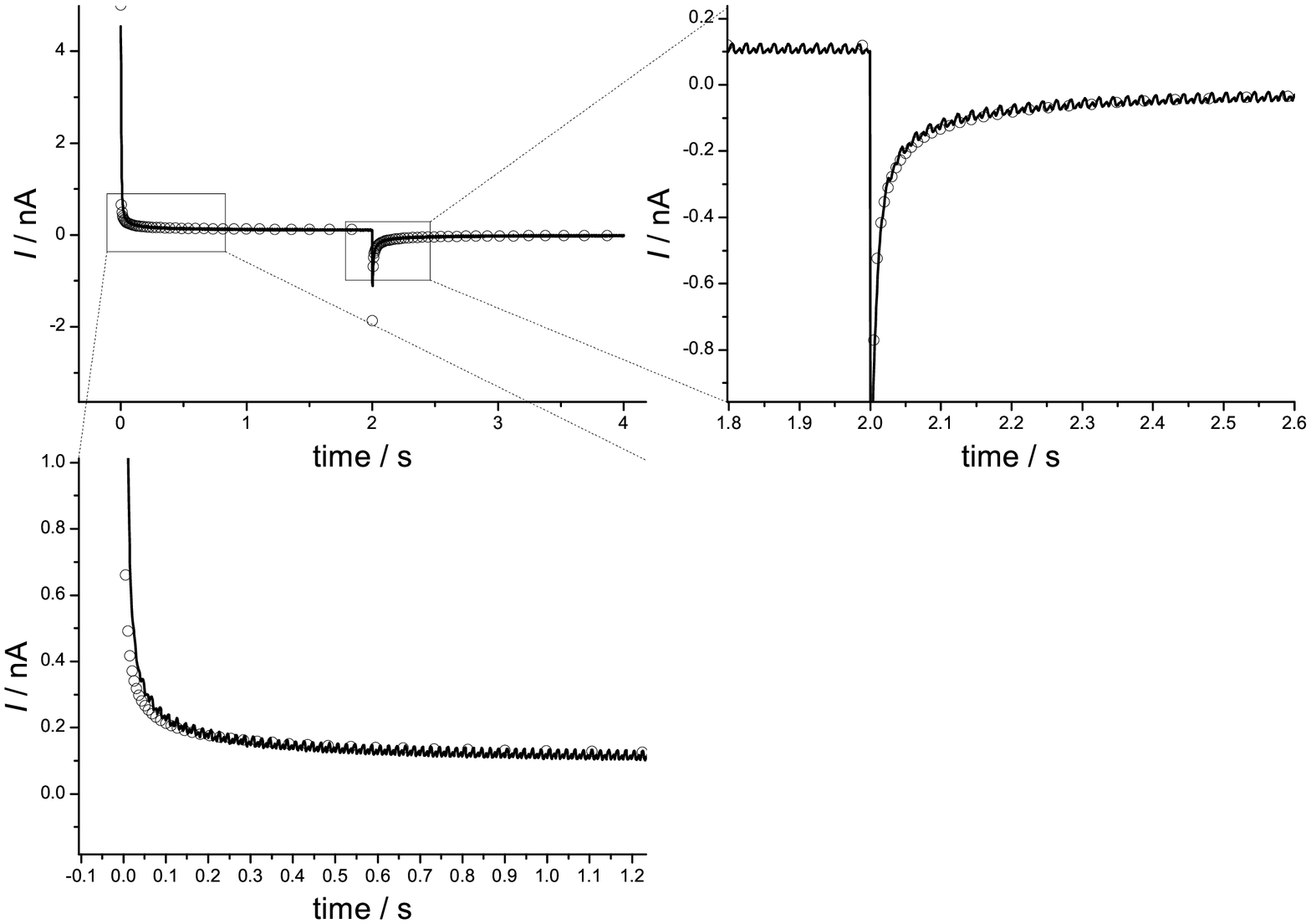}
\caption{Experimental and simulated double potential step chronoamperograms of DMFc at a Pt microdisc electrode, with parts shown zoomed in for clarity.} \label{DMFC DISC}
\end{center}
\end{figure}

Double potential step chronoamperometry was carried out on 2.0 mM decamethylferrocene (DMFc) in the room temperature ionic liquid 1-propyl-3-methylimidazolium bistrifluoromethylsulfonylimide (PmimNTf$_2$) on both a planar microdisc electrode ($r_\mathrm{disc}$ = 5.0 $\mu$m) and the microbandband (width = 50 $\mu$m, length = 6.28 mm) electrode fabricated as described above. In both cases, the potential was initially held at -0.25 V for 20 s, before being stepped to +0.2 V for 2 s then to -0.25 V for 2 s, all vs a silver wire reference electrode. The results for the disc electrode are shown in Figure \ref{DMFC DISC}. Via the simulations, diffusion coefficients for DMFc and DMFc$^+$ of $2.49 (\pm 0.2) \times 10^{-11}$ m$^2$ s$^{-1}$ and $9.57 (\pm 0.6) \times 10^{-12}$ m$^2$ s$^{-1}$ respectively were determined in PmimNTf$_2$. For comparison, a value for the diffusion coefficient of ferrocene in the similar ionic liquid 1-ethyl-3-methylimidazolium bistrifluoromethylsulfonylimide (EmimNTf$_2$) has been found to be $4.7 \times 10 ^{-11}$ m$^2$ s$^{-1}$\cite{Meng2011}. The lower value for DMFc is consistent with this due to its bulkier size.

\begin{figure}[h]
\begin{center}
\includegraphics[width = 0.9\textwidth]{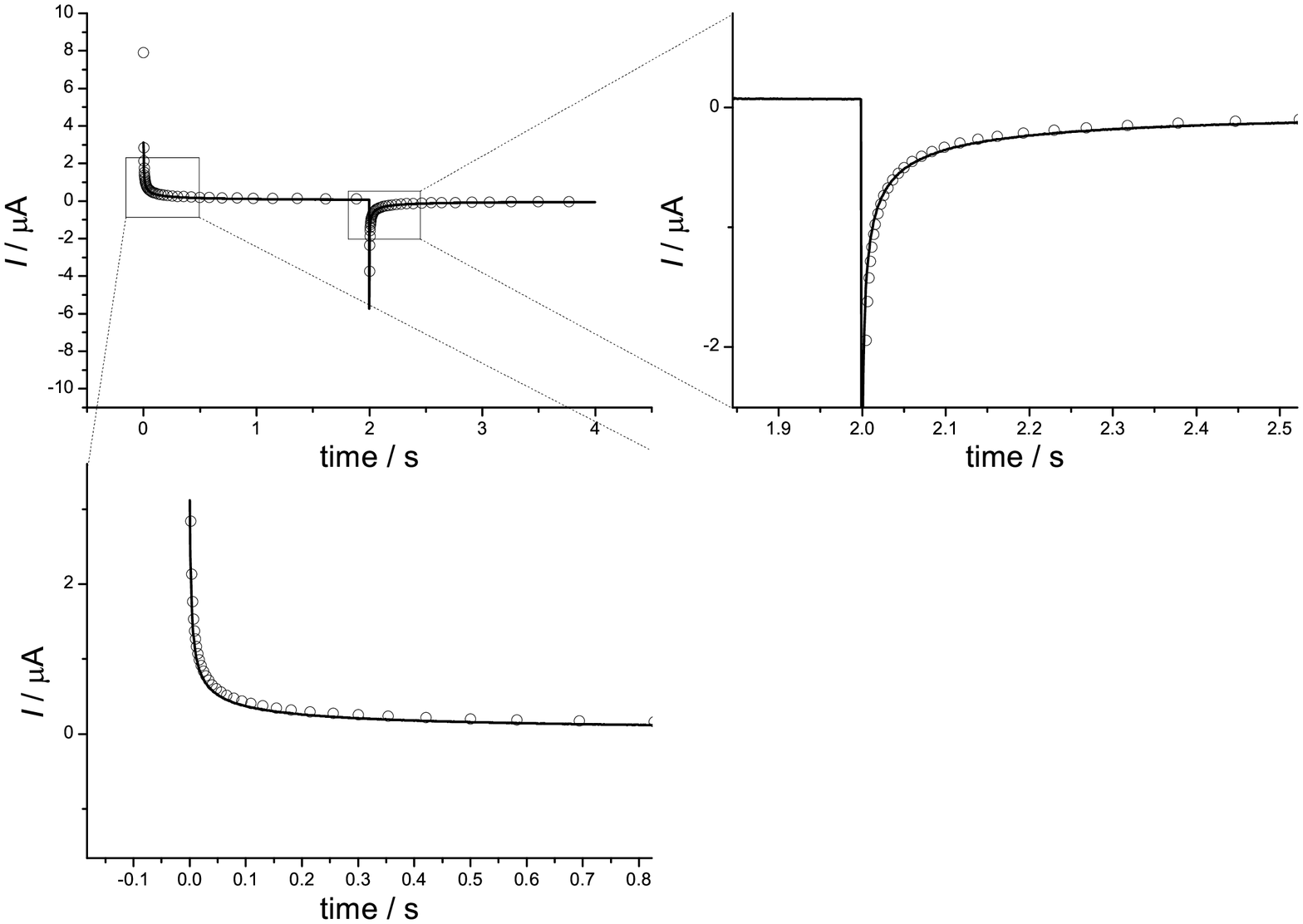}
\caption{Experimental and simulated double potential step chronoamperograms of DMFc at a Pt microband electrode, with parts shown zoomed in for clarity.} \label{DMFC BAND}
\end{center}
\end{figure}

The same experiment was then carried out at a band electrode. The results are shown in Figure \ref{DMFC BAND}. Diffusion coefficients of DMFc and DMFc$^+$ were determined to be $2.50 (\pm 0.2) \times 10^{-11}$ m$^2$ s$^{-1}$ and $9.50 (\pm 0.2) \times 10^{-12}$ m$^2$ s$^{-1}$ respectively in PmimNTf$_2$, in good agreement with those determined from the disc electrode data. This suggests that the program used to simulate the band electrode data is accurate.

\subsection{Cyclic Voltammetry of Decamethylferrocene}

\begin{figure}[h]
\begin{center}
\includegraphics[width = 0.9\textwidth]{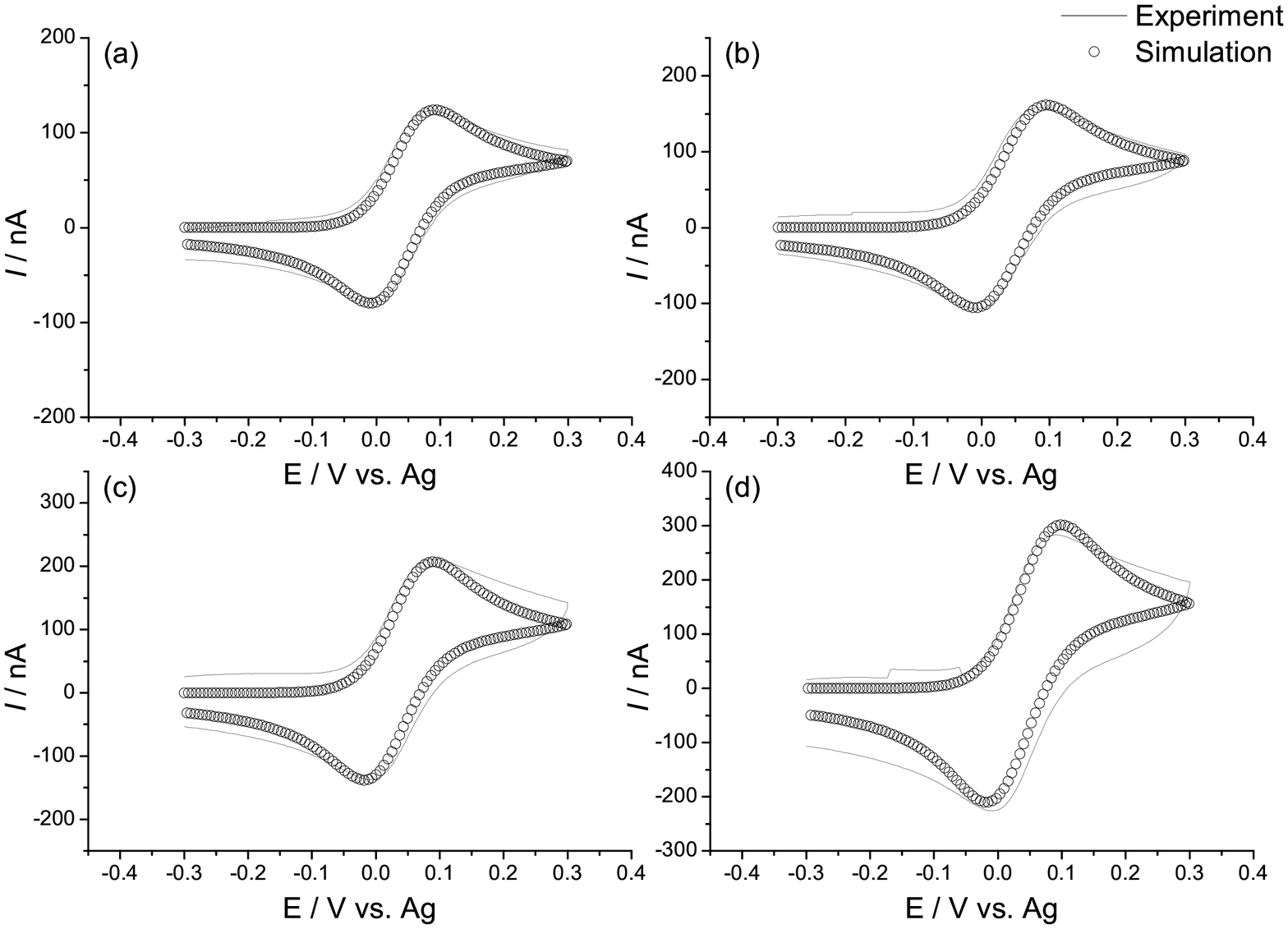}
\caption{Experimental and simulated cyclic voltammograms of DMFc at a Pt microband electrode at scan rates of (a) 50 mV s$^{-1}$, (b) 100 mV s$^{-1}$, (c) 200 mV s$^{-1}$ and (d) 500 mV s$^{-1}$} \label{CV}
\end{center}
\end{figure}

Cyclic voltammetry was carried out on a 2 mM solution of DMFc in the ionic liquid PmimNTF$_2$, at scan rates of 50, 100, 200 and 500 mV s$^{-1}$, using the same microband electrode (width = 50 $\mu$m, length = 6.28 mm). The results, along with simulations, are shown in Figure \ref{CV}. In each case $k^0 = 7 \times 10^{-6}$ m s$^{-1}$, and $\alpha = 0.5$. Good agreement is seen, using the same diffusion coefficients as established for DMFc and DMFc$^{+}$ via double potential step chronoamperometry at the microband electrode.

\section{Conclusions}

A computational model for the simulation of double potential step chronoamperometry at a microband electrode has been developed, which is shown to successfully reproduce experimental data. It has been shown that microband electrodes can usefully be employed to simultaneously determine the diffusion coefficients of both members of a redox couple. Double potential step chronoamperometry was carried out on decamethylferrocene (DMFc) in the ionic liquid 1-propyl-3-methylimidazolium bistrifluoromethylsulfonylimide (PmimNTf$_2$) at both a microdisc electrode and a microband electrode, and it was found that the well established double potential step simulations for a disc produced the same diffusion coefficients as the program developed here for a microband electrode.

\section*{Acknowledgments}

We thank EPSRC (grant no. EP/I029095/1), St John's College, Oxford, and Schlumberger Cambridge Research for funding.

\clearpage

\providecommand*\mcitethebibliography{\thebibliography}
\csname @ifundefined\endcsname{endmcitethebibliography}
  {\let\endmcitethebibliography\endthebibliography}{}

\clearpage

\section*{Figures}

\clearpage

\clearpage

\clearpage

\clearpage

\clearpage

\clearpage

\clearpage

\clearpage

\clearpage

\section*{Tables}

\clearpage

\begin{table}
\begin{center}
\begin{tabular}{l l l}
\hline
Parameter & Description & Units \\
\hline
$\alpha$ & Transfer coefficient & Unitless\\
\\
$c_\mathrm{i}$ & Concentration of species i & mol m$^{-3}$\\
\\
$c_\mathrm{i}^{*}$ & Bulk solution concentration of species i & mol m$^{-3}$\\
\\
$D_\mathrm{i}$ & Diffusion coefficient of species i & m$^2$ s$^{-1}$ \\
\\
$E$ & Applied potential & V\\
\\
$E_f$ & Formal potential of A/B couple & V\\
\\
$F$ & Faraday constant = 96485 & C mol$^{-1}$ \\
\\
$I$ & Current & A\\
\\
$J$ & Flux & mol m$^{-2}$ s$^{-1}$\\
\\
$k^0$ & Electrochemical rate constant & m s$^{-1}$\\ 
\\
$l$ & Electrode length & m \\
\\
$R$ & Gas constant = 8.314 & J K$^{-1}$ mol$^{-1}$\\ 
\\
$r_e$ & Radius of microdisc electrode & m \\
\\
$T$ & Temperature & K\\
\\
$t$ & time & s\\
\\
$w_e$ & width of microband electrode & m \\
\\
$x$ & $x$ coordinate & m \\
\\
$y$ & $y$ coordinate & m \\
\\
$z$ & $z$ coordinate & m \\ 
\hline
\end{tabular}
\end{center}
\caption{List of symbols}
\label{DIMENSIONAL}
\end{table}

\clearpage

\begin{table}
\begin{center}
\begin{tabular}{c c}
Dimensionless Parameter & Definition \\
\hline
\\
$C_\mathrm{i}$ & $\frac{c_\mathrm{i}}{c_\mathrm{A}^*}$ \\
\\
$D_\mathrm{i}^{'}$ & $\frac{D_\mathrm{i}}{D_\mathrm{A}}$ \\
\\
$j$ & $\frac{w_e}{C_\mathrm{A}^{*}D_\mathrm{A}}J$\\
\\
$K^0$ & $\frac{w_e}{D_\mathrm{A}}k^0$\\
\\
$\theta$ & $\frac{RT}{F}\left(E - E_f\right)$\\
\\
$\tau$ & $\frac{D_\mathrm{A}}{w^\mathrm{2}_\mathrm{e}}t$ \\
\\
$X$ & $\frac{x}{w_e}$\\
\\
$Y$ & $\frac{y}{w_e}$\\
\\
\hline
\end{tabular}
\end{center}
\caption{Dimensionless parameters. Species A refers to the species initially present in solution before the experiment/simulation begins.}
\label{DIMENSIONLESS}
\end{table}

\clearpage

\begin{table}
\begin{center}
\begin{tabular}{|c|c|c|}
\hline
\multicolumn{2}{ |c| }{Boundary} & Condition \\
\hline
 & & \\
$\tau < 0$ & all $X$, all $Y$ & $C_\mathrm{A} = 1$, $C_\mathrm{B} = 0$\\
 & & \\
\hline
 & & \\
$0 \leq \tau < \tau_s$ & $0 \leq X \leq 0.5$, $Y = 0$ & $C_\mathrm{A} = 0$, $D^{'}_\mathrm{B}\left(\frac{\partial{C_\mathrm{B}}}{\partial{Y}}\right) = -D^{'}_\mathrm{A}\left(\frac{\partial{C_\mathrm{A}}}{\partial{Y}}\right)$ \\
 & & \\
\hline
 & & \\
$\tau \geq \tau_s$ & $0 \leq X \leq 0.5$, $Y = 0$ & $D^{'}_\mathrm{A}\left(\frac{\partial{C_\mathrm{A}}}{\partial{Y}}\right) = -D^{'}_\mathrm{B}\left(\frac{\partial{C_\mathrm{B}}}{\partial{Y}}\right)$, $C_\mathrm{B} = 0$ \\
 & & \\
\hline
 & & \\
\multirow{10}{*}{All $\tau$} & $X > 0.5$, $Y = 0$ & $\frac{\partial{C_\mathrm{i}}}{\partial{Y}} = 0$ \\
& & \\
\cline{2-3}
& & \\
& $X = 0$, all $Y$ & $\frac{\partial{C_\mathrm{i}}}{\partial{X}} = 0$ \\
& & \\
\cline{2-3}
& & \\
& $X = X_\mathrm{max}$, all $Y$ & \multirow{4}{*}{$C_\mathrm{A} = 1$, $C_\mathrm{B} = 0$} \\
& & \\
\cline{2-2}
& & \\
& All $X$, $Y = Y_\mathrm{max}$ & \\
 & & \\
\hline
\end{tabular}
\end{center}
\caption{Dimensionless boundary conditions for double potential step chronoamperometry simulations}
\label{BOUNDARY CONDITIONS}
\end{table}

\clearpage

\begin{table}
\begin{center}
\begin{tabular}{l l l l l l l l l}
\hline
\multirow{2}{*}{$D_\mathrm{B}^{'}$} & \multicolumn{8}{l}{$\tau - \tau_s$} \\
\cline{2-9}
& 0.01 & 0.1 & 0.5 & 1 & 2 & 5 & 10 & 25 \\
\hline
0.01 & -9.01868 & -2.14050 & -0.56101 & -0.27268 & -0.12090 & -0.03683 & -0.01425 & -0.00395 \\
0.05 & -8.24194 & -1.96945 & -0.52046 & -0.25375 & -0.11277 & -0.03467 & -0.01372 & -0.00401 \\
0.1 & -7.73710 & -1.85671 & -0.49248 & -0.24073 & -0.10781 & -0.03393 & -0.01373 & -0.00413 \\
0.2 & -7.07703 & -1.70899 & -0.45804 & -0.22653 & -0.10341 & -0.03358 & -0.01391 & -0.00430 \\
0.3 & -6.62131 & -1.60827 & -0.43687 & -0.21863 & -0.10123 & -0.03348 & -0.01404 & -0.00440 \\
0.4 & -6.27523 & -1.53294 & -0.42212 & -0.21331 & -0.09979 & -0.03343 & -0.01414 & -0.00447 \\
0.5 & -5.99846 & -1.47360 & -0.41101 & -0.20935 & -0.09871 & -0.03338 & -0.01420 & -0.00453 \\
0.6 & -5.76936 & -1.42515 & -0.40218 & -0.20619 & -0.09783 & -0.03333 & -0.01425 & -0.00457 \\
0.7 & -5.57495 & -1.38458 & -0.39491 & -0.20358 & -0.09709 & -0.03327 & -0.01428 & -0.00460 \\
0.8 & -5.40682 & -1.34991 & -0.38875 & -0.20134 & -0.09644 & -0.03322 & -0.01430 & -0.00462 \\
0.9 & -5.25924 & -1.31981 & -0.38342 & -0.19938 & -0.09586 & -0.03316 & -0.01432 & -0.00464 \\
1.0 & -5.12813 & -1.29334 & -0.37874 & -0.19765 & -0.09534 & -0.03311 & -0.01433 & -0.00466 \\
2.0 & -4.30369 & -1.13273 & -0.34973 & -0.18649 & -0.09175 & -0.03261 & -0.01433 & -0.00474 \\
3.0 & -3.86558 & -1.05096 & -0.33406 & -0.18013 & -0.08953 & -0.03221 & -0.01427 & -0.00477 \\
4.0 & -3.57911 & -0.99829 & -0.32347 & -0.17569 & -0.08791 & -0.03188 & -0.01421 & -0.00478 \\
5.0 & -3.37175 & -0.96029 & -0.31553 & -0.17229 & -0.08664 & -0.03161 & -0.01414 & -0.00478 \\
6.0 & -3.21215 & -0.93096 & -0.30923 & -0.16955 & -0.08560 & -0.03138 & -0.01408 & -0.00477 \\
7.0 & -3.08415 & -0.90729 & -0.30403 & -0.16726 & -0.08471 & -0.03117 & -0.01403 & -0.00477 \\
8.0 & -2.97836 & -0.88757 & -0.29962 & -0.16530 & -0.08394 & -0.03099 & -0.01397 & -0.00476 \\
9.0 & -2.88894 & -0.87077 & -0.29579 & -0.16359 & -0.08326 & -0.03082 & -0.01393 & -0.00476 \\
10 & -2.81198 & -0.85618 & -0.29243 & -0.16207 & -0.08266 & -0.03068 & -0.01388 & -0.00475 \\
20 & -2.37293 & -0.76955 & -0.27154 & -0.15244 & -0.07871 & -0.02965 & -0.01356 & -0.00470 \\
50 & -1.94199 & -0.67568 & -0.24712 & -0.14075 & -0.07369 & -0.02825 & -0.01307 & -0.00460 \\
\hline
\end{tabular}
\end{center}
\caption{Simulated dimensionless fluxes at various times after the second potential step and various values of $D^{'}_\mathrm{B}$ for $\tau_s= 1$}
\label{TAUS=1}
\end{table}

\clearpage

\begin{table}
\begin{center}
\begin{tabular}{l l l l l l l l l}
\hline
\multirow{2}{*}{$D_\mathrm{B}^{'}$} & \multicolumn{8}{l}{$\tau - \tau_s$} \\
\cline{2-9}
& 0.01 & 0.1 & 0.5 & 1 & 2 & 5 & 10 & 25 \\
\hline
0.01 & -16.3990 & -4.74989 & -1.79839 & -1.11825 & -0.65806 & -0.29077 & -0.14069 & -0.04762 \\
0.05 & -12.7921 & -3.73057 & -1.42978 & -0.89715 & -0.53533 & -0.24434 & -0.12297 & -0.04449 \\
0.1 & -10.9238 & -3.20638 & -1.24525 & -0.79048 & -0.48001 & -0.22577 & -0.11639 & -0.04336 \\
0.2 & -9.10659 & -2.70276 & -1.07424 & -0.69448 & -0.43116 & -0.20902 & -0.11009 & -0.04208 \\
0.3 & -8.11739 & -2.43271 & -0.98576 & -0.64532 & -0.40588 & -0.19993 & -0.10648 & -0.04123 \\
0.4 & -7.45936 & -2.25532 & -0.92882 & -0.61359 & -0.38929 & -0.19378 & -0.10395 & -0.04060 \\
0.5 & -6.97637 & -2.12654 & -0.88796 & -0.59064 & -0.37713 & -0.18915 & -0.10201 & -0.04010 \\
0.6 & -6.60022 & -2.02721 & -0.85662 & -0.57291 & -0.36761 & -0.18547 & -0.10044 & -0.03969 \\
0.7 & -6.29536 & -1.94744 & -0.83148 & -0.55859 & -0.35987 & -0.18243 & -0.09913 & -0.03933 \\
0.8 & -6.04110 & -1.88142 & -0.81067 & -0.54665 & -0.35335 & -0.17985 & -0.09800 & -0.03902 \\
0.9 & -5.82440 & -1.82557 & -0.79302 & -0.53646 & -0.34776 & -0.17761 & -0.09701 & -0.03874 \\
1.0 & -5.63655 & -1.77746 & -0.77777 & -0.52761 & -0.34287 & -0.17563 & -0.09614 & -0.03849 \\
2.0 & -4.53860 & -1.50237 & -0.68868 & -0.47496 & -0.31323 & -0.16334 & -0.09057 & -0.03685 \\
3.0 & -4.00197 & -1.37121 & -0.64430 & -0.44809 & -0.29774 & -0.15671 & -0.08748 & -0.03589 \\
4.0 & -3.66493 & -1.28935 & -0.61570 & -0.43053 & -0.28749 & -0.15224 & -0.08536 & -0.03522 \\
5.0 & -3.42694 & -1.23144 & -0.59498 & -0.41769 & -0.27992 & -0.14890 & -0.08377 & -0.03471 \\
6.0 & -3.24689 & -1.18739 & -0.57894 & -0.40768 & -0.27399 & -0.14626 & -0.08250 & -0.03430 \\
7.0 & -3.10429 & -1.15223 & -0.56594 & -0.39953 & -0.26913 & -0.14408 & -0.08144 & -0.03395 \\
8.0 & -2.98759 & -1.12323 & -0.55509 & -0.39270 & -0.26504 & -0.14224 & -0.08054 & -0.03365 \\
9.0 & -2.88971 & -1.09869 & -0.54582 & -0.38684 & -0.26152 & -0.14063 & -0.07976 & -0.03339 \\
10 & -2.80602 & -1.07759 & -0.53775 & -0.38172 & -0.25844 & -0.13924 & -0.07907 & -0.03316 \\
20 & -2.33720 & -0.95451 & -0.48951 & -0.35082 & -0.23963 & -0.13056 & -0.07476 & -0.03169 \\
50 & -1.88948 & -0.82607 & -0.43657 & -0.31626 & -0.21821 & -0.12042 & -0.06962 & -0.02988 \\
\hline
\end{tabular}
\end{center}
\caption{Simulated dimensionless fluxes at various times after the second potential step and various values of $D^{'}_\mathrm{B}$ for $\tau_s= 10$}
\label{TAUS=10}
\end{table}

\clearpage

\begin{table}
\begin{center}
\begin{tabular}{l l l l l l l l l}
\hline
\multirow{2}{*}{$D_\mathrm{B}^{'}$} & \multicolumn{8}{l}{$\tau - \tau_s$} \\
\cline{2-9}
& 0.01 & 0.1 & 0.5 & 1 & 2 & 5 & 10 & 25 \\
\hline
0.01 & -25.46857 & -7.86544 & -3.36292 & -2.29762 & -1.54696 & -0.88524 & -0.55949 & -0.28381 \\
0.05 & -16.17973 & -5.04990 & -2.20413 & -1.53060 & -1.05557 & -0.63350 & -0.41942 & -0.22729\\
0.1 & -12.9585 & -4.08646 & -1.81803 & -1.28084 & -0.90015 & -0.55546 & -0.37504 & -0.20796 \\
0.2 & -10.2775 & -3.29439 & -1.50860 & -1.08353 & -0.77761 & -0.49225 & -0.33790 & -0.19100 \\
0.3 & -8.94730 & -2.90645 & -1.36054 & -0.98933 & -0.71834 & -0.46072 & -0.31891 & -0.18204 \\
0.4 & -8.10251 & -2.66256 & -1.26858 & -0.93049 & -0.68084 & -0.44038 & -0.30649 & -0.17609 \\
0.5 & -7.50047 & -2.49022 & -1.20396 & -0.88886 & -0.65403 & -0.42564 & -0.29741 & -0.17168 \\
0.6 & -7.04138 & -2.35978 & -1.15512 & -0.85717 & -0.63346 & -0.41422 & -0.29032 & -0.16821 \\
0.7 & -6.67523 & -2.25643 & -1.11638 & -0.83189 & -0.61693 & -0.40497 & -0.28455 & -0.16537 \\
0.8 & -6.37373 & -2.17183 & -1.08459 & -0.81101 & -0.60321 & -0.39724 & -0.27971 & -0.16297 \\
0.9 & -6.11946 & -2.10082 & -1.05781 & -0.79336 & -0.59155 & -0.39064 & -0.27556 & -0.16090 \\
1.0 & -5.90099 & -2.04016 & -1.03482 & -0.77812 & -0.58145 & -0.38489 & -0.27193 & -0.15909 \\
2.0 & -4.65808 & -1.69822 & -0.90309 & -0.68954 & -0.52193 & -0.35050 & -0.25002 & -0.14798 \\
3.0 & -4.07023 & -1.53997 & -0.83906 & -0.64563 & -0.49191 & -0.33283 & -0.23860 & -0.14209 \\
4.0 & -3.70712 & -1.44220 & -0.79834 & -0.61739 & -0.47243 & -0.32123 & -0.23105 & -0.13816 \\
5.0 & -3.45342 & -1.37483 & -0.76911 & -0.59698 & -0.45826 & -0.31272 & -0.22549 & -0.13524 \\
6.0 & -3.26290 & -1.32280 & -0.74662 & -0.58118 & -0.44724 & -0.30607 & -0.22112 & -0.13294 \\
7.0 & -3.11285 & -1.28145 & -0.72851 & -0.56840 & -0.43829 & -0.30065 & -0.21755 & -0.13105 \\
8.0 & -2.99059 & -1.24634 & -0.71345 & -0.55773 & -0.43079 & -0.29609 & -0.21453 & -0.12945 \\
9.0 & -2.88841 & -1.21875 & -0.70062 & -0.54862 & -0.42438 & -0.29218 & -0.21194 & -0.12807 \\
10 & -2.80130 & -1.19406 & -0.68950 & -0.54070 & -0.41878 & -0.28875 & -0.20967 & -0.12685 \\
20 & -2.31747 & -1.05162 & -0.62364 & -0.49340 & -0.38513 & -0.26797 & -0.19579 & -0.11937 \\
50 & -1.86142 & -0.90484 & -0.55257 & -0.44155 & -0.34771 & -0.24446 & -0.17989 & -0.11067 \\
\hline
\end{tabular}
\end{center}
\caption{Simulated dimensionless fluxes at various times after the second potential step and various values of $D^{'}_\mathrm{B}$ for $\tau_s= 100$}
\label{TAUS=100}
\end{table}

\clearpage

\begin{table}
\begin{center}
\begin{tabular}{l l l l l l l l l}
\hline
\multirow{2}{*}{$D_\mathrm{B}^{'}$} & \multicolumn{8}{l}{$\tau - \tau_s$} \\
\cline{2-9}
& 0.01 & 0.1 & 0.5 & 1 & 2 & 5 & 10 & 25 \\
\hline
0.01 & -32.0709 & -10.1354 & -4.52168 & -3.19135 & -2.25141 & -1.41638 & -0.99696 & -0.62339 \\
0.05 & -18.2934 & -5.87610 & -2.69999 & -1.94733 & -1.41526 & -0.93900 & -0.69253 & -0.45965 \\
0.1 & -14.1992 & -4.62521 & -2.17650 & -1.59591 & -1.18353 & -0.80733 & -0.60642 & -0.41056 \\
0.2 & -10.9836 & -3.65232 & -1.77703 & -1.33012 & -1.00773 & -0.70472 & -0.53750 & -0.36999 \\
0.3 & -9.44556 & -3.19192 & -1.59109 & -1.20622 & -0.92460 & -0.65483 & -0.50332 & -0.34942 \\
0.4 & -8.48790 & -2.90746 & -1.47706 & -1.12972 & -0.87262 & -0.62310 & -0.48134 & -0.33603 \\
0.5 & -7.81419 & -2.70866 & -1.39755 & -1.07597 & -0.83574 & -0.60032 & -0.46544 & -0.32626 \\
0.6 & -7.30526 & -2.55935 & -1.33779 & -1.03529 & -0.80762 & -0.58279 & -0.45314 & -0.31869 \\
0.7 & -6.90233 & -2.44175 & -1.29058 & -1.00296 & -0.78512 & -0.56867 & -0.44319 & -0.31248 \\
0.8 & -6.57253 & -2.34592 & -1.25197 & -0.97637 & -0.76652 & -0.55692 & -0.43489 & -0.30730 \\
0.9 & -6.29576 & -2.26582 & -1.21954 & -0.95393 & -0.75076 & -0.54692 & -0.42779 & -0.30286 \\
1.0 & -6.05896 & -2.19753 & -1.19175 & -0.93462 & -0.73714 & -0.53825 & -0.42162 & -0.29898 \\
2.0 & -4.72941 & -1.81818 & -1.03370 & -0.82321 & -0.65759 & -0.48689 & -0.38478 & -0.27562 \\
3.0 & -4.11100 & -1.64323 & -0.95759 & -0.76853 & -0.61792 & -0.46083 & -0.36587 & -0.26347 \\
4.0 & -3.73234 & -1.53577 & -0.90944 & -0.73357 & -0.59233 & -0.44384 & -0.35347 & -0.25544 \\
5.0 & -3.46926 & -1.46053 & -0.87500 & -0.70839 & -0.57379 & -0.43145 & -0.34438 & -0.24953 \\
6.0 & -3.27250 & -1.40373 & -0.84857 & -0.68896 & -0.55941 & -0.42180 & -0.33727 & -0.24488 \\
7.0 & -3.11801 & -1.35867 & -0.82732 & -0.67327 & -0.54776 & -0.41394 & -0.33147 & -0.24108 \\
8.0 & -2.99244 & -1.32167 & -0.80969 & -0.66020 & -0.53803 & -0.40736 & -0.32660 & -0.23788 \\
9.0 & -2.88770 & -1.29049 & -0.79469 & -0.64906 & -0.52971 & -0.40171 & -0.32242 & -0.23512 \\
10 & -2.79855 & -1.26369 & -0.78170 & -0.63938 & -0.52247 & -0.39678 & -0.31876 & -0.23270 \\
20 & -2.30575 & -1.10962 & -0.70508 & -0.58182 & -0.47908 & -0.36701 & -0.29652 & -0.21792 \\
50 & -1.84474 & -0.95188 & -0.62297 & -0.51918 & -0.43121 & -0.33363 & -0.27132 & -0.20094 \\
\hline
\end{tabular}
\end{center}
\caption{Simulated dimensionless fluxes at various times after the second potential step and various values of $D^{'}_\mathrm{B}$ for $\tau_s= 1000$}
\label{TAUS=1000}
\end{table}


\begin{mcitethebibliography}{55}
\providecommand*\natexlab[1]{#1}
\providecommand*\mciteSetBstSublistMode[1]{}
\providecommand*\mciteSetBstMaxWidthForm[2]{}
\providecommand*\mciteBstWouldAddEndPuncttrue
  {\def\EndOfBibitem{\unskip.}}
\providecommand*\mciteBstWouldAddEndPunctfalse
  {\let\EndOfBibitem\relax}
\providecommand*\mciteSetBstMidEndSepPunct[3]{}
\providecommand*\mciteSetBstSublistLabelBeginEnd[3]{}
\providecommand*\EndOfBibitem{}
\mciteSetBstSublistMode{f}
\mciteSetBstMaxWidthForm{subitem}{(\alph{mcitesubitemcount})}
\mciteSetBstSublistLabelBeginEnd
  {\mcitemaxwidthsubitemform\space}
  {\relax}
  {\relax}

\bibitem[Compton and Banks(2010)Compton, and Banks]{Compton2010}
Compton,~R.~G.; Banks,~C.~E. \emph{Understanding Voltammetry}, 2nd ed.; World
  Scientific: Singapore, 2010\relax
\mciteBstWouldAddEndPuncttrue
\mciteSetBstMidEndSepPunct{\mcitedefaultmidpunct}
{\mcitedefaultendpunct}{\mcitedefaultseppunct}\relax
\EndOfBibitem
\bibitem[Aoki and Osteryoung(1981)Aoki, and Osteryoung]{Aoki1981}
Aoki,~K.; Osteryoung,~J. \emph{J. Electroanal. Chem.} \textbf{1981},
  \emph{122}, 19--35\relax
\mciteBstWouldAddEndPuncttrue
\mciteSetBstMidEndSepPunct{\mcitedefaultmidpunct}
{\mcitedefaultendpunct}{\mcitedefaultseppunct}\relax
\EndOfBibitem
\bibitem[Amatore et~al.(2008)Amatore, Oleinick, and Svir]{Amatore2008a}
Amatore,~C.; Oleinick,~A.; Svir,~I. \emph{Anal. Chem.} \textbf{2008},
  \emph{80}, 7947--7956\relax
\mciteBstWouldAddEndPuncttrue
\mciteSetBstMidEndSepPunct{\mcitedefaultmidpunct}
{\mcitedefaultendpunct}{\mcitedefaultseppunct}\relax
\EndOfBibitem
\bibitem[Barnes et~al.(2010)Barnes, O'Mahony, Aldous, Hardacre, and
  Compton]{Barnes2010a}
Barnes,~E.~O.; O'Mahony,~A.~M.; Aldous,~L.; Hardacre,~C.; Compton,~R.~G.
  \emph{J. Electroanal. Chem.} \textbf{2010}, \emph{646}, 11--17\relax
\mciteBstWouldAddEndPuncttrue
\mciteSetBstMidEndSepPunct{\mcitedefaultmidpunct}
{\mcitedefaultendpunct}{\mcitedefaultseppunct}\relax
\EndOfBibitem
\bibitem[Oldham(1981)]{Oldham1981}
Oldham,~K.~B. \emph{J. Electroanal. Chem.} \textbf{1981}, \emph{122},
  1--17\relax
\mciteBstWouldAddEndPuncttrue
\mciteSetBstMidEndSepPunct{\mcitedefaultmidpunct}
{\mcitedefaultendpunct}{\mcitedefaultseppunct}\relax
\EndOfBibitem
\bibitem[Heinze(1981)]{Heinze1981}
Heinze,~J. \emph{J. Electroanal. Chem.} \textbf{1981}, \emph{124}, 73--86\relax
\mciteBstWouldAddEndPuncttrue
\mciteSetBstMidEndSepPunct{\mcitedefaultmidpunct}
{\mcitedefaultendpunct}{\mcitedefaultseppunct}\relax
\EndOfBibitem
\bibitem[Paddon et~al.(2007)Paddon, Silvester, Bhatti, Donohoe, and
  Compton]{Paddon2007}
Paddon,~C.~A.; Silvester,~D.~S.; Bhatti,~F.~L.; Donohoe,~T.~J.; Compton,~R.~G.
  \emph{Electroanalysis} \textbf{2007}, \emph{19}, 11--22\relax
\mciteBstWouldAddEndPuncttrue
\mciteSetBstMidEndSepPunct{\mcitedefaultmidpunct}
{\mcitedefaultendpunct}{\mcitedefaultseppunct}\relax
\EndOfBibitem
\bibitem[Shoup and Szabo(1982)Shoup, and Szabo]{Shoup1982}
Shoup,~D.; Szabo,~A. \emph{J. Electroanal. Chem.} \textbf{1982}, \emph{140},
  237--245\relax
\mciteBstWouldAddEndPuncttrue
\mciteSetBstMidEndSepPunct{\mcitedefaultmidpunct}
{\mcitedefaultendpunct}{\mcitedefaultseppunct}\relax
\EndOfBibitem
\bibitem[Klymenko et~al.(2004)Klymenko, Evans, Hardacre, Svir, and
  Compton]{Klymenko2004}
Klymenko,~O.~V.; Evans,~R.~G.; Hardacre,~C.; Svir,~I.~B.; Compton,~R.~G.
  \emph{J. Electroanal. Chem.} \textbf{2004}, \emph{571}, 211--221\relax
\mciteBstWouldAddEndPuncttrue
\mciteSetBstMidEndSepPunct{\mcitedefaultmidpunct}
{\mcitedefaultendpunct}{\mcitedefaultseppunct}\relax
\EndOfBibitem
\bibitem[Streeter and Compton(2008)Streeter, and Compton]{Streeter2008b}
Streeter,~I.; Compton,~R.~G. \emph{J. Phys. Chem. C} \textbf{2008}, \emph{112},
  13716--13728\relax
\mciteBstWouldAddEndPuncttrue
\mciteSetBstMidEndSepPunct{\mcitedefaultmidpunct}
{\mcitedefaultendpunct}{\mcitedefaultseppunct}\relax
\EndOfBibitem
\bibitem[Barnes et~al.(2012)Barnes, Lewis, Dale, Marken, and
  Compton]{Barnes2012}
Barnes,~E.~O.; Lewis,~G. E.~M.; Dale,~S. E.~C.; Marken,~F.; Compton,~R.~G.
  \emph{Analyst} \textbf{2012}, \emph{137}, 1068--1081\relax
\mciteBstWouldAddEndPuncttrue
\mciteSetBstMidEndSepPunct{\mcitedefaultmidpunct}
{\mcitedefaultendpunct}{\mcitedefaultseppunct}\relax
\EndOfBibitem
\bibitem[Bai et~al.(2013)Bai, Del~Campo, and Tsai]{Bai2013}
Bai,~Y.-H.; Del~Campo,~J.~F.; Tsai,~Y.-C. \emph{Biosens. Bioelectron.}
  \textbf{2013}, \emph{42}, 17--22\relax
\mciteBstWouldAddEndPuncttrue
\mciteSetBstMidEndSepPunct{\mcitedefaultmidpunct}
{\mcitedefaultendpunct}{\mcitedefaultseppunct}\relax
\EndOfBibitem
\bibitem[Huan et~al.(2012)Huan, Hung, Ha, Anh, Khai, Shim, and Chung]{Huan2012}
Huan,~T.~N.; Hung,~L.~Q.; Ha,~V. T.~T.; Anh,~N.~H.; Khai,~T.~V.; Shim,~K.~B.;
  Chung,~H. \emph{Talanta} \textbf{2012}, \emph{94}, 284 -- 288\relax
\mciteBstWouldAddEndPuncttrue
\mciteSetBstMidEndSepPunct{\mcitedefaultmidpunct}
{\mcitedefaultendpunct}{\mcitedefaultseppunct}\relax
\EndOfBibitem
\bibitem[Metters et~al.(2012)Metters, Kadara, and Banks]{Metters2012}
Metters,~J.~P.; Kadara,~R.~O.; Banks,~C.~E. \emph{Sens. Actuators, B}
  \textbf{2012}, \emph{169}, 136\relax
\mciteBstWouldAddEndPuncttrue
\mciteSetBstMidEndSepPunct{\mcitedefaultmidpunct}
{\mcitedefaultendpunct}{\mcitedefaultseppunct}\relax
\EndOfBibitem
\bibitem[Walsh et~al.(2010)Walsh, Lovelock, and Licence]{Walsh2010}
Walsh,~D.~A.; Lovelock,~K. R.~J.; Licence,~P. \emph{Chem. Soc. Re v.}
  \textbf{2010}, \emph{39}, 4185--4194\relax
\mciteBstWouldAddEndPuncttrue
\mciteSetBstMidEndSepPunct{\mcitedefaultmidpunct}
{\mcitedefaultendpunct}{\mcitedefaultseppunct}\relax
\EndOfBibitem
\bibitem[Broder et~al.(2007)Broder, Silvester, Aldous, Hardacre, Crossley, and
  Compton]{Broder2007}
Broder,~T.~L.; Silvester,~D.~S.; Aldous,~L.; Hardacre,~C.; Crossley,~A.;
  Compton,~R.~G. \emph{New J. Chem.} \textbf{2007}, \emph{31}, 966--972\relax
\mciteBstWouldAddEndPuncttrue
\mciteSetBstMidEndSepPunct{\mcitedefaultmidpunct}
{\mcitedefaultendpunct}{\mcitedefaultseppunct}\relax
\EndOfBibitem
\bibitem[Galvez and Alcaraz(1992)Galvez, and Alcaraz]{Galvez1992}
Galvez,~J.; Alcaraz,~M.-L. \emph{J. Electroanal. Chem.} \textbf{1992},
  \emph{341}, 15--34\relax
\mciteBstWouldAddEndPuncttrue
\mciteSetBstMidEndSepPunct{\mcitedefaultmidpunct}
{\mcitedefaultendpunct}{\mcitedefaultseppunct}\relax
\EndOfBibitem
\bibitem[Limon-Petersen et~al.(2009)Limon-Petersen, Streeter, Rees, and
  Compton]{Limon-Petersen2009}
Limon-Petersen,~J.~G.; Streeter,~I.; Rees,~N.~V.; Compton,~R.~G. \emph{J. Phys.
  Chem. C} \textbf{2009}, \emph{113}, 333--337\relax
\mciteBstWouldAddEndPuncttrue
\mciteSetBstMidEndSepPunct{\mcitedefaultmidpunct}
{\mcitedefaultendpunct}{\mcitedefaultseppunct}\relax
\EndOfBibitem
\bibitem[nan()]{nanoflux}
http://www.nanoflux.com.sg/\relax
\mciteBstWouldAddEndPuncttrue
\mciteSetBstMidEndSepPunct{\mcitedefaultmidpunct}
{\mcitedefaultendpunct}{\mcitedefaultseppunct}\relax
\EndOfBibitem
\bibitem[mic()]{micrux}
http://www.micruxfluidic.com/\relax
\mciteBstWouldAddEndPuncttrue
\mciteSetBstMidEndSepPunct{\mcitedefaultmidpunct}
{\mcitedefaultendpunct}{\mcitedefaultseppunct}\relax
\EndOfBibitem
\bibitem[Welford et~al.(2001)Welford, Freeman, Wilkins, Wadhawan, Hahn, and
  Compton]{Welford2001}
Welford,~P.~J.; Freeman,~J.; Wilkins,~S.~J.; Wadhawan,~J.~D.; Hahn,~C. E.~W.;
  Compton,~R.~G. \emph{Anal. Chem.} \textbf{2001}, \emph{73}, 6088--6092\relax
\mciteBstWouldAddEndPuncttrue
\mciteSetBstMidEndSepPunct{\mcitedefaultmidpunct}
{\mcitedefaultendpunct}{\mcitedefaultseppunct}\relax
\EndOfBibitem
\bibitem[Compton et~al.(1993)Compton, Fisher, Wellington, Dobson, and
  Leigh]{Compton1993b}
Compton,~R.~G.; Fisher,~A.~C.; Wellington,~R.~G.; Dobson,~P.~J.; Leigh,~P.~A.
  \emph{J. Phys. Chem.} \textbf{1993}, \emph{97}, 10410--10415\relax
\mciteBstWouldAddEndPuncttrue
\mciteSetBstMidEndSepPunct{\mcitedefaultmidpunct}
{\mcitedefaultendpunct}{\mcitedefaultseppunct}\relax
\EndOfBibitem
\bibitem[Alden et~al.(1995)Alden, Compton, and Dryfe]{Alden1995}
Alden,~J.~A.; Compton,~R.~G.; Dryfe,~R.~A. \emph{J. Electr} \textbf{1995},
  \emph{397}, 11--17\relax
\mciteBstWouldAddEndPuncttrue
\mciteSetBstMidEndSepPunct{\mcitedefaultmidpunct}
{\mcitedefaultendpunct}{\mcitedefaultseppunct}\relax
\EndOfBibitem
\bibitem[Alden and Compton(1996)Alden, and Compton]{Alden1996a}
Alden,~J.~A.; Compton,~R.~G. \emph{J. Electroanal. Chem.} \textbf{1996},
  \emph{404}, 27 -- 35\relax
\mciteBstWouldAddEndPuncttrue
\mciteSetBstMidEndSepPunct{\mcitedefaultmidpunct}
{\mcitedefaultendpunct}{\mcitedefaultseppunct}\relax
\EndOfBibitem
\bibitem[Alden and Compton(1996)Alden, and Compton]{Alden1996b}
Alden,~J.~A.; Compton,~R.~G. \emph{J. Elec} \textbf{1996}, \emph{402},
  1--10\relax
\mciteBstWouldAddEndPuncttrue
\mciteSetBstMidEndSepPunct{\mcitedefaultmidpunct}
{\mcitedefaultendpunct}{\mcitedefaultseppunct}\relax
\EndOfBibitem
\bibitem[Ueno et~al.(2003)Ueno, Kim, and Kitamura]{Ueno2003}
Ueno,~K.; Kim,~H.-B.; Kitamura,~N. \emph{Anal. Chem.} \textbf{2003}, \emph{75},
  2086--2091\relax
\mciteBstWouldAddEndPuncttrue
\mciteSetBstMidEndSepPunct{\mcitedefaultmidpunct}
{\mcitedefaultendpunct}{\mcitedefaultseppunct}\relax
\EndOfBibitem
\bibitem[Amatore et~al.(2011)Amatore, Lemmer, Sella, and Thouin]{Amatore2011a}
Amatore,~C.; Lemmer,~C.; Sella,~C.; Thouin,~L. \emph{Anal. Chem.}
  \textbf{2011}, \emph{83}, 4170--4177\relax
\mciteBstWouldAddEndPuncttrue
\mciteSetBstMidEndSepPunct{\mcitedefaultmidpunct}
{\mcitedefaultendpunct}{\mcitedefaultseppunct}\relax
\EndOfBibitem
\bibitem[Amatore et~al.(2011)Amatore, Lemmer, Perrodin, Sella, and
  Thouin]{Amatore2011}
Amatore,~C.; Lemmer,~C.; Perrodin,~P.; Sella,~C.; Thouin,~L. \emph{Electrochem.
  Commun.} \textbf{2011}, \emph{13}, 1459 -- 1461\relax
\mciteBstWouldAddEndPuncttrue
\mciteSetBstMidEndSepPunct{\mcitedefaultmidpunct}
{\mcitedefaultendpunct}{\mcitedefaultseppunct}\relax
\EndOfBibitem
\bibitem[Matthews et~al.(2012)Matthews, Shiddiky, Yunus, Elton, Duffy, Gu,
  Fisher, and Bond]{Matthews2012}
Matthews,~S.~M.; Shiddiky,~M. J.~A.; Yunus,~K.; Elton,~D.~M.; Duffy,~N.~W.;
  Gu,~Y.; Fisher,~A.~C.; Bond,~A.~M. \emph{Anal. Chem.} \textbf{2012},
  \emph{84}, 6686--6692\relax
\mciteBstWouldAddEndPuncttrue
\mciteSetBstMidEndSepPunct{\mcitedefaultmidpunct}
{\mcitedefaultendpunct}{\mcitedefaultseppunct}\relax
\EndOfBibitem
\bibitem[Fosset et~al.(1991)Fosset, Amatore, Bartelt, Michael, and
  Wightman]{Fosset1991}
Fosset,~B.; Amatore,~C.~A.; Bartelt,~J.~E.; Michael,~A.~C.; Wightman,~R.~M.
  \emph{Anal. Chem.} \textbf{1991}, \emph{63}, 306--314\relax
\mciteBstWouldAddEndPuncttrue
\mciteSetBstMidEndSepPunct{\mcitedefaultmidpunct}
{\mcitedefaultendpunct}{\mcitedefaultseppunct}\relax
\EndOfBibitem
\bibitem[Unwin(1991)]{Unwin1991}
Unwin,~P.~R. \emph{J. Electroanal. Chem.} \textbf{1991}, \emph{297}, 103 --
  124\relax
\mciteBstWouldAddEndPuncttrue
\mciteSetBstMidEndSepPunct{\mcitedefaultmidpunct}
{\mcitedefaultendpunct}{\mcitedefaultseppunct}\relax
\EndOfBibitem
\bibitem[Rajantie and Williams(2001)Rajantie, and Williams]{Rajantie2001}
Rajantie,~H.; Williams,~D.~E. \emph{Analyst} \textbf{2001}, \emph{126},
  1882--1887\relax
\mciteBstWouldAddEndPuncttrue
\mciteSetBstMidEndSepPunct{\mcitedefaultmidpunct}
{\mcitedefaultendpunct}{\mcitedefaultseppunct}\relax
\EndOfBibitem
\bibitem[Paix{\~a}o et~al.(2003)Paix{\~a}o, Matos, and Bertotti]{Paixao2003}
Paix{\~a}o,~T.~R.; Matos,~R.~C.; Bertotti,~M. \emph{Electrochim. Acta}
  \textbf{2003}, \emph{48}, 691 -- 698\relax
\mciteBstWouldAddEndPuncttrue
\mciteSetBstMidEndSepPunct{\mcitedefaultmidpunct}
{\mcitedefaultendpunct}{\mcitedefaultseppunct}\relax
\EndOfBibitem
\bibitem[Svir et~al.(2003)Svir, Oleinick, and Compton]{Svir2003}
Svir,~I.~B.; Oleinick,~A.~I.; Compton,~R.~G. \emph{J. Electroanal. Chem.}
  \textbf{2003}, \emph{560}, 117--126\relax
\mciteBstWouldAddEndPuncttrue
\mciteSetBstMidEndSepPunct{\mcitedefaultmidpunct}
{\mcitedefaultendpunct}{\mcitedefaultseppunct}\relax
\EndOfBibitem
\bibitem[Amatore et~al.(2006)Amatore, Sella, and Thouin]{Amatore2006}
Amatore,~C.; Sella,~C.; Thouin,~L. \emph{J. Electroanal. Chem.} \textbf{2006},
  \emph{593}, 194 -- 202\relax
\mciteBstWouldAddEndPuncttrue
\mciteSetBstMidEndSepPunct{\mcitedefaultmidpunct}
{\mcitedefaultendpunct}{\mcitedefaultseppunct}\relax
\EndOfBibitem
\bibitem[Compton and Winkler(1995)Compton, and Winkler]{Compton1995a}
Compton,~R.~G.; Winkler,~J. \emph{J. Phys. Chem.} \textbf{1995}, \emph{99},
  5029--5034\relax
\mciteBstWouldAddEndPuncttrue
\mciteSetBstMidEndSepPunct{\mcitedefaultmidpunct}
{\mcitedefaultendpunct}{\mcitedefaultseppunct}\relax
\EndOfBibitem
\bibitem[Alden and Compton(1996)Alden, and Compton]{Alden1996c}
Alden,~J.~A.; Compton,~R.~G. \emph{Electroanalysis} \textbf{1996}, \emph{8},
  30--33\relax
\mciteBstWouldAddEndPuncttrue
\mciteSetBstMidEndSepPunct{\mcitedefaultmidpunct}
{\mcitedefaultendpunct}{\mcitedefaultseppunct}\relax
\EndOfBibitem
\bibitem[Aoki et~al.(1987)Aoki, Tokuda, and Matsuda]{Aoki1987}
Aoki,~K.; Tokuda,~K.; Matsuda,~H. \emph{J. Ele} \textbf{1987}, \emph{230},
  61\relax
\mciteBstWouldAddEndPuncttrue
\mciteSetBstMidEndSepPunct{\mcitedefaultmidpunct}
{\mcitedefaultendpunct}{\mcitedefaultseppunct}\relax
\EndOfBibitem
\bibitem[Barrosse-Antle et~al.(2010)Barrosse-Antle, Bond, Compton, O'Mahony,
  Rogers, and Silvester]{Barrosse-Antle2010}
Barrosse-Antle,~L.~E.; Bond,~A.~M.; Compton,~R.~G.; O'Mahony,~A.~M.;
  Rogers,~E.~I.; Silvester,~D.~S. \emph{Chem. Asian J.} \textbf{2010},
  \emph{5}, 202--230\relax
\mciteBstWouldAddEndPuncttrue
\mciteSetBstMidEndSepPunct{\mcitedefaultmidpunct}
{\mcitedefaultendpunct}{\mcitedefaultseppunct}\relax
\EndOfBibitem
\bibitem[Silvester and Compton(2006)Silvester, and Compton]{Silvester2006a}
Silvester,~D.~S.; Compton,~R.~G. \emph{Z. Phys. Chem.} \textbf{2006},
  \emph{220}, 1247--1274\relax
\mciteBstWouldAddEndPuncttrue
\mciteSetBstMidEndSepPunct{\mcitedefaultmidpunct}
{\mcitedefaultendpunct}{\mcitedefaultseppunct}\relax
\EndOfBibitem
\bibitem[Xiong et~al.(2012)Xiong, Fletcher, Ernst, Davies, and
  Compton]{Xiong2012}
Xiong,~L.; Fletcher,~A.~M.; Ernst,~S.; Davies,~S.~G.; Compton,~R.~G.
  \emph{Analyst} \textbf{2012}, \emph{137}, 2567\relax
\mciteBstWouldAddEndPuncttrue
\mciteSetBstMidEndSepPunct{\mcitedefaultmidpunct}
{\mcitedefaultendpunct}{\mcitedefaultseppunct}\relax
\EndOfBibitem
\bibitem[Xiong et~al.(2012)Xiong, Fletcher, Davies, Norman, Hardacre, and
  Compton]{Xiong2012a}
Xiong,~L.; Fletcher,~A.~M.; Davies,~S.~G.; Norman,~S.~E.; Hardacre,~C.;
  Compton,~R.~G. \emph{Analyst} \textbf{2012}, \emph{137}, 4951\relax
\mciteBstWouldAddEndPuncttrue
\mciteSetBstMidEndSepPunct{\mcitedefaultmidpunct}
{\mcitedefaultendpunct}{\mcitedefaultseppunct}\relax
\EndOfBibitem
\bibitem[Buzzeo et~al.(2003)Buzzeo, Klymenko, Wadhawan, Hardacre, Seddon, and
  Compton]{Buzzeo2003}
Buzzeo,~M.~C.; Klymenko,~O.~V.; Wadhawan,~J.~D.; Hardacre,~C.; Seddon,~K.~R.;
  Compton,~R.~G. \emph{J. Phys. Chem. A} \textbf{2003}, \emph{107},
  8872--8878\relax
\mciteBstWouldAddEndPuncttrue
\mciteSetBstMidEndSepPunct{\mcitedefaultmidpunct}
{\mcitedefaultendpunct}{\mcitedefaultseppunct}\relax
\EndOfBibitem
\bibitem[Gavaghan(1998)]{Gavaghan1998b}
Gavaghan,~D.~J. \emph{J. Electroanal. Chem.} \textbf{1998}, \emph{456},
  25--35\relax
\mciteBstWouldAddEndPuncttrue
\mciteSetBstMidEndSepPunct{\mcitedefaultmidpunct}
{\mcitedefaultendpunct}{\mcitedefaultseppunct}\relax
\EndOfBibitem
\bibitem[Gavaghan(1998)]{Gavaghan1998a}
Gavaghan,~D.~J. \emph{J. Electroanal. Chem.} \textbf{1998}, \emph{456},
  13--23\relax
\mciteBstWouldAddEndPuncttrue
\mciteSetBstMidEndSepPunct{\mcitedefaultmidpunct}
{\mcitedefaultendpunct}{\mcitedefaultseppunct}\relax
\EndOfBibitem
\bibitem[Gavaghan(1998)]{Gavaghan1998}
Gavaghan,~D.~J. \emph{J. Electroanal. Chem.} \textbf{1998}, \emph{456},
  1--12\relax
\mciteBstWouldAddEndPuncttrue
\mciteSetBstMidEndSepPunct{\mcitedefaultmidpunct}
{\mcitedefaultendpunct}{\mcitedefaultseppunct}\relax
\EndOfBibitem
\bibitem[Crank and Nicolson(1947)Crank, and Nicolson]{Crank1947}
Crank,~J.; Nicolson,~E. \emph{Proc. Camb. Phil. Soc.} \textbf{1947}, \emph{43},
  50--67\relax
\mciteBstWouldAddEndPuncttrue
\mciteSetBstMidEndSepPunct{\mcitedefaultmidpunct}
{\mcitedefaultendpunct}{\mcitedefaultseppunct}\relax
\EndOfBibitem
\bibitem[Press et~al.(2007)Press, Teukolsky, Vetterling, and
  Flannery]{Press2007}
Press,~W.~H., Teukolsky,~S.~A., Vetterling,~W.~T., Flannery,~B.~P., Eds.
  \emph{Numerical Recipes: The Art of Scientific Computing}; Cambridge
  University Press, 2007\relax
\mciteBstWouldAddEndPuncttrue
\mciteSetBstMidEndSepPunct{\mcitedefaultmidpunct}
{\mcitedefaultendpunct}{\mcitedefaultseppunct}\relax
\EndOfBibitem
\bibitem[Amatore and Svir(2003)Amatore, and Svir]{Amatore2003}
Amatore,~C.; Svir,~I. \emph{Journal of Electroanalytical Chemistry}
  \textbf{2003}, \emph{557}, 75 -- 90\relax
\mciteBstWouldAddEndPuncttrue
\mciteSetBstMidEndSepPunct{\mcitedefaultmidpunct}
{\mcitedefaultendpunct}{\mcitedefaultseppunct}\relax
\EndOfBibitem
\bibitem[Amatore et~al.(2004)Amatore, Oleinick, and Svir]{Amatore2004}
Amatore,~C.; Oleinick,~A.; Svir,~I. \emph{Electrochem. Commun.} \textbf{2004},
  \emph{6}, 1123 -- 1130\relax
\mciteBstWouldAddEndPuncttrue
\mciteSetBstMidEndSepPunct{\mcitedefaultmidpunct}
{\mcitedefaultendpunct}{\mcitedefaultseppunct}\relax
\EndOfBibitem
\bibitem[Barnes et~al.(2013)Barnes, Zhou, Rees, and Compton]{Barnes2013}
Barnes,~E.~O.; Zhou,~Y.-G.; Rees,~N.~V.; Compton,~R.~G. \emph{Journal of
  Electroanalytical Chemistry} \textbf{2013}, \emph{691}, 28 -- 34\relax
\mciteBstWouldAddEndPuncttrue
\mciteSetBstMidEndSepPunct{\mcitedefaultmidpunct}
{\mcitedefaultendpunct}{\mcitedefaultseppunct}\relax
\EndOfBibitem
\bibitem[Rogers et~al.(2008)Rogers, Silvester, Poole, Aldous, Hardacre, and
  Compton]{Rogers2008a}
Rogers,~E.~I.; Silvester,~D.~S.; Poole,~D.~L.; Aldous,~L.; Hardacre,~C.;
  Compton,~R.~G. \emph{J. Phys. Chem. C} \textbf{2008}, \emph{112},
  2729--2735\relax
\mciteBstWouldAddEndPuncttrue
\mciteSetBstMidEndSepPunct{\mcitedefaultmidpunct}
{\mcitedefaultendpunct}{\mcitedefaultseppunct}\relax
\EndOfBibitem
\bibitem[Evans et~al.(2004)Evans, Klymenko, Saddoughi, Hardacre, and
  Compton]{Evans2004}
Evans,~R.~G.; Klymenko,~O.~V.; Saddoughi,~S.~A.; Hardacre,~C.; Compton,~R.~G.
  \emph{J. Phys. Chem. B} \textbf{2004}, \emph{108}, 7878--7886\relax
\mciteBstWouldAddEndPuncttrue
\mciteSetBstMidEndSepPunct{\mcitedefaultmidpunct}
{\mcitedefaultendpunct}{\mcitedefaultseppunct}\relax
\EndOfBibitem
\bibitem[Meng et~al.(2011)Meng, Aldous, and Compton]{Meng2011}
Meng,~Y.; Aldous,~L.; Compton,~R.~G. \emph{J. Phys. Chem. C} \textbf{2011},
  \emph{115}, 14334--14340\relax
\mciteBstWouldAddEndPuncttrue
\mciteSetBstMidEndSepPunct{\mcitedefaultmidpunct}
{\mcitedefaultendpunct}{\mcitedefaultseppunct}\relax
\EndOfBibitem
\end{mcitethebibliography}
\end{document}